\documentclass[aps,preprint,showpacs,amsmath,amssymb,superscriptaddress]{revtex4}
\usepackage{graphicx}

\begin{document}
\newcommand {\nc} {\newcommand}
\nc {\beq} {\begin{eqnarray}}
\nc {\eol} {\nonumber \\}
\nc {\eeq} {\end{eqnarray}}
\nc {\eeqn} [1] {\label{#1} \end{eqnarray}}
\nc {\eoln} [1] {\label{#1} \\}
\nc {\ve} [1] {\mbox{\boldmath $#1$}}
\nc {\vS} {\mbox{$\ve{S}$}}
\nc {\cA} {\mbox{$\cal{A}$}}
\nc {\dem} {\mbox{$\frac{1}{2}$}}
\nc {\arrow} [2] {\mbox{$\mathop{\rightarrow}\limits_{#1 \rightarrow #2}$}}

\title{Analysis of the $^6$He $\beta$ decay into the $\alpha+d$ continuum within a three-body model}

\author{E.M. Tursunov}
\email{tursune@inp.uz}
\affiliation{Institute of Nuclear Physics, Uzbekistan Academy of Sciences, 702132, Ulugbek, Tashkent, Uzbekistan}
\affiliation{Physique Nucl\'eaire Th\'eorique et Physique Math\'ematique, C.P. 229,
Universit\'e Libre de Bruxelles, B 1050 Brussels, Belgium}
\author{D. Baye}  
\email{dbaye@ulb.ac.be}
\affiliation{Physique Quantique, C.P. 165/82,  
Universit\'e Libre de Bruxelles, B 1050 Brussels, Belgium}
\affiliation{Physique Nucl\'eaire Th\'eorique et Physique Math\'ematique, C.P. 229,
Universit\'e Libre de Bruxelles, B 1050 Brussels, Belgium}
\author{P. Descouvemont}
\email{pdesc@ulb.ac.be}
\affiliation{Physique Nucl\'eaire Th\'eorique et Physique Math\'ematique, C.P. 229,
Universit\'e Libre de Bruxelles, B 1050 Brussels, Belgium}


\date{\today}
\begin{abstract}
The $\beta$-decay process of the $^6$He halo nucleus into the $\alpha+d$ continuum 
is studied in a three-body model. 
The $^6$He nucleus is described as an $\alpha+n+n$ system in hyperspherical coordinates 
on a Lagrange mesh. 
The convergence of the Gamow-Teller matrix element requires the knowledge 
of wave functions up to about 30 fm and of hypermomentum components up to $K=24$. 
The shape and absolute values of the transition probability per time and energy units 
of a recent experiment can be reproduced very well with an appropriate $\alpha+d$ potential. 
A total transition probability of $1.6 \times 10^{-6}$ s$^{-1}$ is obtained 
in agreement with that experiment. 
Halo effects are shown to be very important because of a strong cancellation 
between the internal and halo components of the matrix element, as observed in previous studies. 
The forbidden bound state in the $\alpha+d$ potential is found essential 
to reproduce the order of magnitude of the data. 
Comments are made on $R$-matrix fits. 
\end{abstract}
\maketitle
%
\section{Introduction}
The discovery of light halo nuclei with very large matter radii near the neutron drip line 
\cite{tan96} inspired detailed studies of the structure of these quantum systems. 
The large radii are interpreted as arising from an extended spatial density of a few neutrons 
\cite{han87,zhu91}. 
The static properties of the halo nuclei do not provide a complete picture of their structure 
and especially of the halo extension. 
Few observables directly probe the probability density at very large distances. 

The $\beta$ decay with emission of a deuteron, also known as $\beta$ delayed deuteron 
decay, is energetically possible for nuclei with a two-neutron separation energy $S_{2n}$ 
smaller than 3 MeV. 
This property is typical of halo nuclei. 
The measurement of the spectrum shape for this decay process offers a unique opportunity of probing 
halo properties at large distances. 
The $\beta$ decay of $^6$He into $\alpha$ and a deuteron has been observed in several 
experiments \cite{rii90,bor93,ABB02}. 
The branching ratio is much smaller than expected from simple $R$-matrix \cite{rii90}, 
two-body \cite{Pierre92}, and three-body \cite{zhu93} models. 
Various experimental values of this branching ratio have been obtained, 
i.e., $(2.8 \pm 0.5) \times 10^{-6}$ \cite{rii90}, $(7.6 \pm 0.6) \times 10^{-6}$ \cite{bor93}, 
and $(1.9 \pm 0.8) \times 10^{-6}$ \cite{ABB02}, for a deuteron cutoff energy of about 350 keV. 
The latest result \cite{ABB02} is a factor of 4 smaller than the result of Ref.~\cite{bor93} 
which served as reference for most theoretical papers. 

A semi-microscopic study \cite{BSD94} of the process has been able to explain 
that the low value of the branching ratio is the result of a cancellation 
between the "internal" and "external" parts of the Gamow-Teller matrix element. 
The overlaps of the $^6$He ground state and $\alpha+d$ scattering wave functions 
in the internal ($R<5$ fm) and external ($R>5$ fm) regions have very close 
magnitudes but opposite signs. 
It is clear that the external part of the Gamow-Teller matrix element reflects properties 
of the halo structure of the $^6$He nucleus. 
An improved microscopic wave function of $^6$He confirmed this interpretation \cite{VSO94}. 
It was also confirmed by a fit in the $R$-matrix framework \cite{bar94} 
which yields a satisfactory description of the deuteron spectrum shape and branching ratio 
of Ref.~\cite{bor93}. 
A fully microscopic description of the $\beta$ decay of the $^6$He nucleus to the $^6$Li ground state 
and to the $\alpha+d$ continuum \cite{cso94} was performed in a dynamical microscopic cluster model 
with consistent fully antisymmetrized wave functions for the initial bound state 
and the final scattering state. 
This model provided a reasonable agreement with the data of Ref.~\cite{bor93}. 
Without any fitted parameter, those data were underestimated by about a factor of 2. 
Hence, the same microscopic results now {\it overestimate} the recent data of Ref.~\cite{ABB02} 
by a similar factor. 

Since new data \cite{ABB02} with much better statistics 
which provide an even lower branching ratio are now available, 
it is timely to reexamine the interpretation of the $\beta$ delayed deuteron decay. 
Improving significantly the microscopic model of Ref.~\cite{cso94} is not yet possible. 
We prefer thus to base our discussion on an $\alpha+n+n$ three-body model. 
Accurate wave functions of $^6$He are available in hyperspherical coordinates \cite{DDB03}. 
A previous calculation based on the same model \cite{zhu93} contains several limitations 
which led to a significant overestimation of the data of Ref.~\cite{bor93}: 
the calculations were restricted to small values of the hypermomentum, $K=0$ and 2, 
and the halo description may not have been sufficiently extended. 

The aim of the present work is the determination of the deuteron spectrum shape and branching ratio 
for the $\beta$ decay of the $^6$He halo nucleus into $\alpha+d$ 
with an accurate treatment of the $^6$He wave function in an $\alpha+n+n$ three-body cluster model. 
For the description of the structure of the $^6$He nucleus, we use the hyperspherical harmonics 
method on a Lagrange mesh \cite{BH-86,DDB03} which yields an accurate solution in this model. 
The $\alpha+d$ scattering wave function is treated as factorized into a deuteron 
wave function and a nucleus-nucleus relative wave function. 
We will choose several versions of the central interaction potential between $\alpha$ and $d$: 
a deep Gaussian potential \cite{dub94} which fits both the $s$-wave phase shift 
and the binding energy of the $^6$Li ground state (1.473 MeV), 
and potentials obtained by folding the $\alpha+N$ potential of Voronchev et al. \cite{vor95}. 
For the sake of comparison we will also perform a calculation with a repulsive $\alpha+d$ potential 
which was used in Ref.~\cite{zhu93}. 

In Sec.~II, the formalism of the $\beta$-decay process of the $^6$He nucleus into the $\alpha+d$ continuum 
is presented. 
The potentials and the corresponding three-body hyperspherical and two-body scattering wave functions 
are also described. 
In Sec.~III, we discuss the obtained numerical results in comparison with experimental data. 
Finally conclusions are given in Sec.~IV.        
     
\section{Model}

\subsection{$^6$He wave function}

The initial three-body wave function is expressed in hyperspherical coordinates. 
Particles 1 and 2 are the nucleons and 3 is the $\alpha$ cluster. 
A set of Jacobi coordinates for the three particles with mass numbers $A_1=1$, $A_2=1$, and $A_3=4$ 
is defined as 
\beq
\ve{x} = \sqrt{\mu_{12}} \ve{r}, \hspace*{1 cm}
\ve{y} = \sqrt{\mu_{(12)3}} \ve{R},
\label{eq201}
\eeq
where $\ve{r} = \ve{r}_2-\ve{r}_1$ and $\ve{R} = \ve{r}_3 - \dem (\ve{r}_1+\ve{r}_2)$. 
The (dimensionless) reduced masses are given by $\mu_{12}=1/2$ and $\mu_{(12)3}=4/3$. 
Equations (\ref{eq201}) define six coordinates which are transformed to the hyperspherical coordinates by 
\beq
\rho^2 = x^2+y^2, \hspace*{1 cm} 
\alpha = \arctan (y/x),
\label{eq203}
\eeq
where $\alpha$ varies between 0 and $\pi/2$. 
With the angular variables $\Omega_x = (\theta_x,\varphi_x)$ and $\Omega_y =(\theta_y,\varphi_y)$, 
Eqs.~(\ref{eq203}) define a set of hyperspherical coordinates. 
These coordinates are known to be well adapted to the three-body Schr{\"o}dinger equation. 

With the notation $\Omega_5 = (\alpha, \Omega_x, \Omega_y)$, the wave function reads \cite{DDB03} 
\beq
\Psi^{00+}_{^6{\rm He}} (\rho,\Omega_5) 
= \rho^{-5/2} \sum_{l_x l_y LS K} {\chi}^{0+}_{l_x l_y LS K}(\rho) 
{\cal Y}^{00}_{l_x l_y LS K}
(\Omega_5),
\label{eq210}
\eeq
where $l_x$ and $l_y$ are the orbital momenta
associated with the Jacobi coordinates $\ve{x}$ and $\ve{y}$, respectively, 
${\cal Y}^{JM}_{l_x l_y LS K}$ are hyperspherical harmonics, 
and ${\chi}^{J\pi}_{l_x l_y LS K}$ are hyperradial functions. 

The $^6$He ground state wave function contains components with total intrinsic spin $S=0$ and 1. 
The total orbital momentum $L$ is equal to $S$. 
Because of the positive parity, $l_x+l_y$ is even and 
the sums in Eq.~(\ref{eq210}) and in the following run over even $K$ values only. 

\subsection{$\alpha+d$ wave function}

For the scattering state, we assume an expression factorized into the deuteron ground-state wave function 
and an $\alpha+d$ scattering wave function derived from a potential model. 
The deuteron spin 1 and positive parity allow $S$ and $D$ components. 
Here, we neglect the small $D$ component of the deuteron. 

Below, we only need the $1^+$ component of the $\alpha+d$ scattering function which reads 
\beq
\Psi^{1M+}_{\alpha d}(E,\ve{r},\ve{R}) = \Psi_d(\ve{r}) \psi_{\alpha d}(E,\ve{R}).
\label{eq215}
\eeq
where $\ve{r}$ and $\ve{R}$ are here the deuteron and $\alpha+d$ relative coordinates, respectively. 
The spatial part of the deuteron wave function $\Psi_d$ is written as 
\beq
\psi_d(\ve{r}) = r^{-1} u_d(r) Y_{00}(\Omega_x).
\label{eq216}
\eeq
The spatial part of the $\alpha+d$ $s$-wave function is factorized as 
\beq
\psi_{\alpha d}(E,\ve{R}) =  R^{-1} u_E(R) Y_{l0}(\Omega_y). 
\label{eq217}
\eeq
The radial scattering wave function $u_E$ has the asymptotic behavior,
\beq
u_E(R) \arrow{R}{\infty} \cos\delta_0(E) F_0 (kR) + \sin\delta_0(E) G_0(kR),
\label{eq220}
\eeq 
where $k$ is the wave number of the relative motion, $F_0$ and $G_0$ are Coulomb functions, 
and $\delta_0(E)$ is the $s$-wave phase shift at energy $E$. 

The radial functions $u_E$ are calculated with effective $\alpha+d$ potentials. 
Some among the potentials $V_{\alpha d}(R)$ we are using are obtained by folding 
an $\alpha+N$ potential $V_{\alpha N}(r)$. 
They are given by the equation 
\beq
V_{\alpha d}(R) = \langle \psi_d(\ve{r}) \mid V_{\alpha n}(|\ve{R}+\dem\ve{r}|) 
+ V_{\alpha p}(|\ve{R}-\dem\ve{r}|) \mid \psi_d(\ve{r}) \rangle,
\label{eq301}
\eeq
where the integration is performed over the radial and angular parts of variable $\ve{r}$.

\subsection{Transition probability per time and energy units}

For the $\beta$-decay process 
\begin{equation}
^6{\rm He} \to \alpha + d + {\rm e}^- + \bar{\nu},
\end{equation}     
the transition probability per time and energy units is given by \cite{BD-88}
\begin{equation}
\frac{dW}{dE}= \frac{m_ec^2}{\pi^4 v \hbar^2} G_{\beta}^2 f(Q-E) B_{\rm GT}(E),
\label{eq310}
\end{equation}
where $m_e$ is the electron mass, $v$ and $E$ are the relative velocity and energy in the center 
of mass system of $\alpha$ and deuteron, and $G_{\beta}=2.996 \times 10^{-12}$ is the dimensionless 
$\beta$-decay constant \cite{wil82}. 
The Fermi integral $f(Q-E)$ depends on the kinetic energy $Q-E$, available for the electron and antineutrino. 
The mass difference $Q$ between initial and final particles is 2.03 MeV. 

Between an initial state with isospin $T=1$ and a final state with isospin $T=0$, 
Gamow-Teller transitions are allowed. 
The reduced transition probability reads
\begin{equation}
B_{\rm GT}(E) = 12 \lambda ^2 \sum_M |\langle \Psi^{1M+}_{\alpha d}(E)|\sum_{j=1}^2 t_{j-} s_{jz}| 
\Psi^{00+}_{^6{\rm He}} \rangle|^2,
\label{eq300}
\end{equation} 
where $\lambda=1.268 \pm 0.002$ is the ratio of the axial-vector to vector coupling constants \cite{dub90}, 
$\Psi^{1M+}_{\alpha d}(E)$ is the wave function (\ref{eq215}) of the final $\alpha+d$ system, 
and $\Psi^{00+}_{^6{\rm He}}$ is the wave function (\ref{eq210}) of $^6$He. 
The operators $\ve{s}_j$ and $\ve{t}_j$ are the spin and isospin operators of particle $j$, respectively. 

Since the total orbital momentum and parity are conserved, only the $l=0$ partial scattering wave contributes. 
Hence, only the initial $L=S=0$ component of $^6$He can decay to $\alpha+d$. 
It is convenient to express the Gamow-Teller matrix element 
with the help of an effective wave function \cite{BSD94}, 
\beq
\psi_{\rm eff}(\ve{R}) = \int \psi_d(\ve{r}) \psi^0_{^6{\rm He}}(\ve{r},\ve{R}) d\ve{r}.
\label{eq101}
\eeq 
In this expression, $\psi^0_{^6{\rm He}}(\ve{r},\ve{R})$ is the spatial part 
of the $S=0$ component of the $^6$He wave function. 
The reduced transition probability can be written as 
\begin{equation}
B_{\rm GT}(E) = 6\lambda ^2 \left[ \int \psi_{\alpha d}(E,\ve{R}) \psi_{\rm eff}(\ve{R}) d\ve{R} \right]^2.
\label{eq102}
\end{equation} 

Since only $l_x=l_y=L=S=0$ contributes, let us define 
\begin{equation}
Z_K(r,R) = \rho^{-5/2} \chi_{0000K} (\rho) \mathcal{N}_K P_{K/2}^{1/2, 1/2}(\cos 2\alpha),  
\end{equation}
where $\mathcal{N}_K$ is a normalisation factor coming from ${\cal Y}^{00}_{0000K}$ 
given by Eq.~(10) of Ref.~\cite{DDB03} and $P_{K/2}^{1/2, 1/2}$ is a Jacobi polynomial \cite{abr70}. 
After integration over all angles, the reduced transition probability (\ref{eq102}) becomes 
\beq
B_{\rm GT}(E)= 6\lambda^2 \left[ \sum_{K} \int_0^{\infty} u_E(R) u^{(K)}_{\rm eff}(R) dR \right]^2.
\label{eq104}
\eeq
It involves the $K$-dependent effective functions 
\beq
u^{(K)}_{\rm eff}(R)= R \int_0^{\infty} Z_{K}(r,R) u_d(r) r dr, 
\label{eq105}
\eeq
the sum of which forms the radial part of $\psi_{\rm eff}(\ve{R})$. 

\section{Results and discussion}

\subsection{Conditions of the calculation}

The central Minnesota interaction \cite{thom77} reproduces the deuteron binding energy 
and fairly approximates the low-energy nucleon-nucleon scattering properties. 
The deuteron wave function $\psi_d$ is calculated over a Lagrange-Laguerre mesh involving 40 mesh points 
and a scaling parameter $h=0.25$ fm (see Ref.~\cite{DDB03} for details). 
An energy $E_d = -2.202$ MeV is obtained. 
The calculations are done with $\hbar^2/2 m_N=20.734$ MeV fm$^2$. 

The initial $\alpha+n+n$ bound state is calculated as explained in Ref.~\cite{DDB03}. 
The number of components is limited to $K_{\rm max} = 24$. 
The same nucleon-nucleon interaction is used, i.e., the Minnesota interaction 
with an exchange parameter $u = 1$. 
The $\alpha+N$ potential is however different from the one employed in Ref.~\cite{DDB03}. 
Here we employ the $\alpha+N$ potential of Voronchev et al.\ \cite{vor95} 
with a multiplicative factor 1.035 in order to reproduce the $^6$He binding energy. 
This change of interaction is motivated by the fact that we want to use the same 
interaction for the derivation of the $\alpha+d$ folding potential. 
The renormalization factor slightly affects the $^5$He properties.
For the $3/2^-$ resonance, the original potential of Ref.~\cite{vor95} provides 
an energy $E_R=0.80$ MeV and a width $\Gamma=0.75$ MeV, in nice agreement with experiment. 
Introducing the renormalization factor provides $E_R=0.55$ MeV and $\Gamma=0.40$ MeV, 
but does not affect the unstable nature of the resonance.

Since the valence neutron and proton in the $^6$Li nucleus belong to the $0p_{3/2}$ subshell, 
we use the $p$-wave $\alpha+N$ potential of Ref.~\cite{vor95} when deriving 
the $\alpha+d$ folding potential by using Eq.~(\ref{eq301}). 
For the $s$ wave, this potential yields two bound states for $^6$Li 
with energies $E_0=-19.87$ MeV and $E_1=-0.83$ MeV, respectively. 
The first one is forbidden by the Pauli principle and the second one is underbound 
compared with the experimental ground-state energy $E_{\rm exp}=-1.473$ MeV. 
The $\alpha+N$ potential of Kanada et al.\ \cite{KKN79} employed in Ref.~\cite{DDB03} does not yield 
an $\alpha+d$ folding potential with a physical bound state in the $s$ wave. 

The numerical calculations of the Gamow-Teller matrix elements are done with 
several potentials. 
(i) The simple Gaussian attractive potential $V_a = -76.12 \exp(- 0.2 r^2)$ \cite{dub94} 
simultaneously provides the correct $^6$Li binding energy 
(together with a forbidden state) and a fair fit of the low-energy experimental phase shifts. 
(ii) The folding potential does not reproduce the $^6$Li ground-state energy. 
Therefore, we multiply the central part of the original $\alpha+N$ potential by a factor $f_1=1.068$. 
The $\alpha+d$ folding potential $V_{f1}$ moves the physical state to $E_1=-1.470$ MeV. 
(iii) The folding potential $V_{f1}$ does not have the same quality of phase shift description 
as the simple Gaussian potential $V_a$. 
Therefore, we also choose another multiplicative factor $f_2=1.15$ for the folding potential $V_{f2}$, 
which gives a stronger binding for the $^6$Li ground state, $E_1=-2.386$ MeV. 
(iv) Finally, for the sake of comparison with Ref.~\cite{zhu93}, we also 
perform a calculation with their Woods-Saxon repulsive potential ($V_r$). 
Of course, it does not bind $^6$Li. 

\begin{figure}[thb]
\begin{center}
\includegraphics[width=10cm]{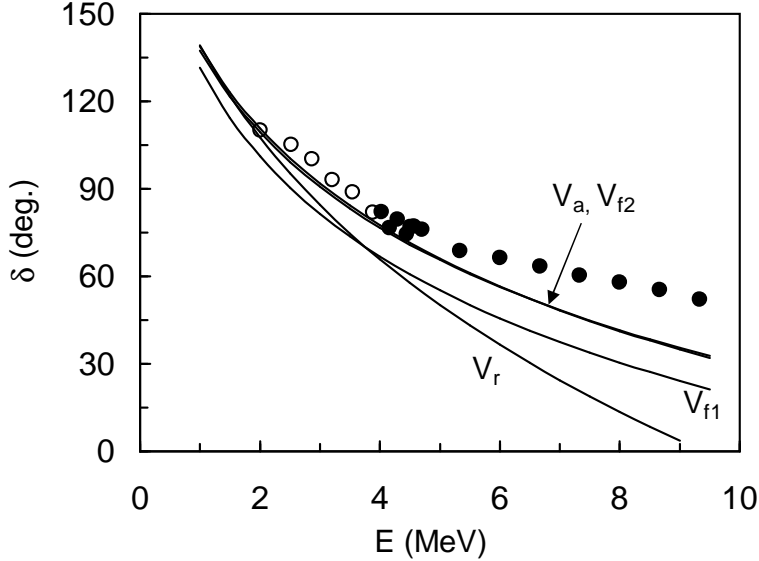}
\end{center}
\caption{$s$-wave phase shifts obtained with different $\alpha+d$ potentials: 
attractive Gaussian potential $V_a$ \cite{dub94}, 
folding potentials $V_{f1}$ and $V_{f2}$ (see text), 
and repulsive Woods-Saxon potential $V_r$ \cite{zhu93}. 
Phase shifts are taken from analyses of experimental data in Refs.~\cite{GSK75} (open dots) 
and \cite{JGK83} (full dots). 
\label{Fig1}}
\end{figure}

The $s$-wave phase shifts for the different $\alpha+d$ potentials are compared in Fig.~\ref{Fig1} 
with the results of phase-shift analyses \cite{GSK75,JGK83}. 
The description of the $\alpha+d$ phase shift is poorest for the repulsive potential. 
Fair, almost identical, phase shifts are obtained with $V_a$ and $V_{f2}$. 

In Eqs.~(\ref{eq105}) and (\ref{eq301}), the integration over $r$ is done by using 
the Gauss-Laguerre quadrature consistent with the Lagrange mesh. 
This ensures numerical convergence for the transition probability.
The integration over variable $R$ in Eq.~(\ref{eq104}) is performed with the simple trapezoidal rule 
with a step 0.05 fm. 
Later we show that with this choice of step, 
convergent results for the transition probability are obtained with 600 points, 
which corresponds to a maximal $\alpha+d$ relative distance $R_{\rm max}=30$ fm. 

We have also calculated the Gamow-Teller matrix element for the $\beta$ decay to the $^6$Li ground state. 
For this calculation, we replace wave function $\Psi^{1+}_{\alpha d}(E)$ in Eq.~(\ref{eq300}) 
by the $\alpha+n+p$ wave function of the $^6$Li ground state 
obtained with the same nuclear interactions as for $^6$He. 
The result $B_{\rm GT}=4.489 \lambda^2$ is about 5 \% below the experimental value 
$B_{\rm GT}^{(\rm exp)}=4.745 \lambda^2$. 
With the potential of Ref.~\cite{KKN79} for the $^6$He description, we find $B_{\rm GT}=4.636 \lambda^2$. 
The sensitivity with respect to the $^6$He wave function is therefore small.

\subsection{Effective wave functions and their integrals}

\begin{figure}[b]
\begin{center}
\includegraphics[width=10cm]{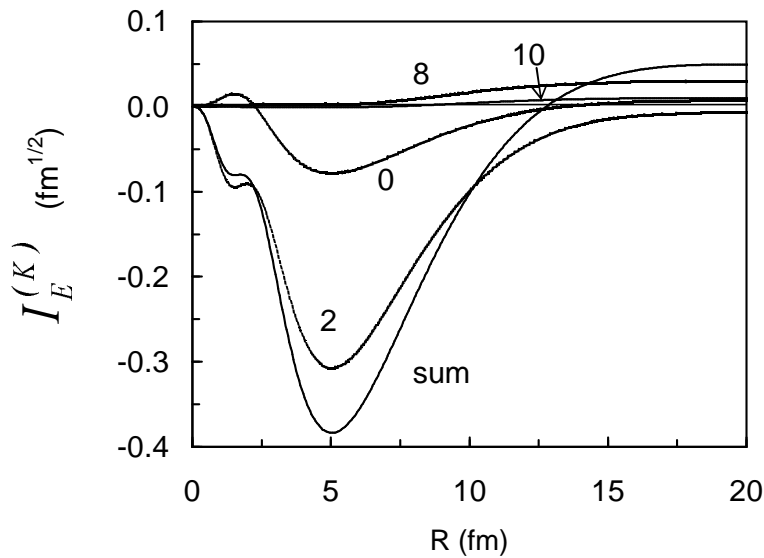}
\end{center}
\caption{Integrals $I_E^{(K)}(R)$ [Eq.~\ref{eq302})] at the energy $E=1$ MeV 
for the $\alpha+d$ potential of Ref.~\cite{dub94} and different $K$ values 
with $K_{\rm max}=24$ and $R_{\rm max}=30$ fm. 
The other components would hardly be visible at the scale of the figure. 
\label{Fig2}}
\end{figure}

In order to analyze the cancellation effects in the Gamow-Teller matrix element 
for the $\beta$ delayed deuteron decay, we display in Fig.~\ref{Fig2} the integrals 
\beq
I_E^{(K)}(R) = \int_0^R u_E(R') u_{\rm eff}^{(K)}(R') dR'
\label{eq302}
\eeq
at the $\alpha+d$ relative energy $E=1$ MeV for different $K$ values. 
They are calculated by using the $\alpha+d$ potential of Ref.~\cite{dub94}. 
The reduced transition probability is given by the limit 
\beq
B_{\rm GT}(E) = 6\lambda^2 \left[ \lim_{R\to\infty} \sum_K I_E^{(K)}(R) \right]^2.  
\label{eq304}
\eeq
>From Fig.~\ref{Fig2}, one can see that at large $R$ values the dominant contribution 
to $\sum_K I_E^{(K)}$ for all $K$ values up to $K_{\rm max}$ 
comes from the $K=2$ and $K=8$ components. 
Components for $K=4$ and $K=6$ as well as for $K \ge 12$ are not visible 
with the linear scale of Fig.~\ref{Fig2}. 
Although the $K=0$ component is rather important around $R = 5$ fm, 
it is suppressed at large $R$ values even more than the $K=10$ component. 

\begin{figure}[hbt]
\begin{center}
\includegraphics[width=10cm]{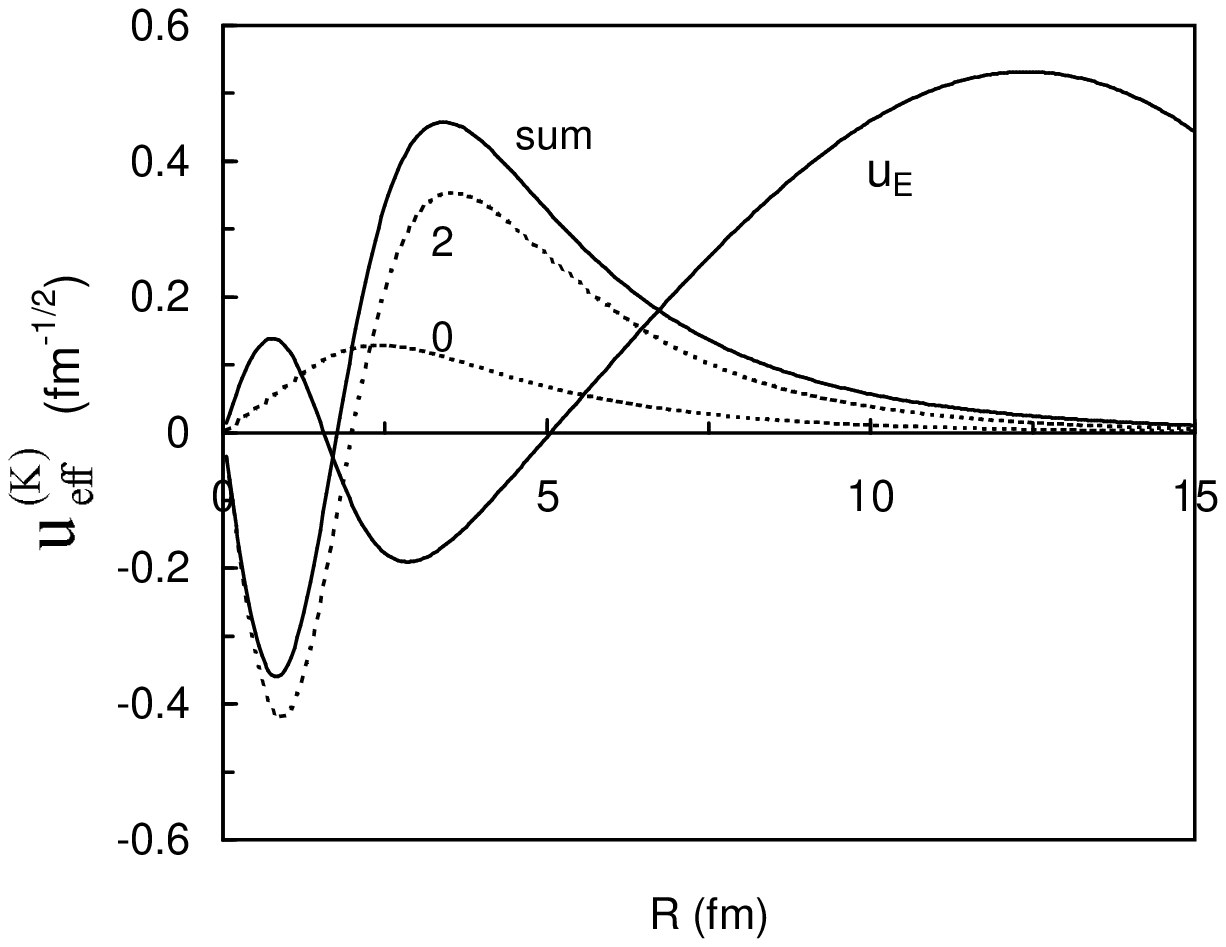}
\end{center}
\caption{Effective wave functions $u_{\rm eff}^{(K)}(R)$ [Eq.~(\ref{eq105})] 
and $u_{\rm eff}(R)$ [Eq.~(\ref{eq305})] for the $\alpha+d$ potential of Ref.~\cite{dub94} 
and different $K$ values with $K_{\rm max}=24$ and $R_{\rm max}=30$ fm. 
The scattering wave function $u_E$ at 1 MeV is also represented.
\label{Fig3}}
\end{figure}

To understand this interesting effect, we display different components $u_{\rm eff}^{(K)}(R)$ 
of the effective wave function as dotted lines in Fig.~\ref{Fig3}. 
The full line represents the sum
\beq
u_{\rm eff}(R) = \sum_{K=0}^{K_{\rm max}} u_{\rm eff}^{(K)}(R).
\eeqn{eq305}
In Fig.~\ref{Fig3}, we also show the $\alpha+d$ scattering $s$-wave function $u_E$ for $E=1$ MeV. 
It is important to note that this function keeps a constant sign in the interval 5-19 fm. 
This constant-sign interval is even broader for smaller values of $E$. 
The $K=0$ and $K=2$ components are dominant at all relative distances $R$. 
They exhibit a maximum below 5 fm. 
One observes that $u_{\rm eff}^{(0)}$ keeps a constant sign over the whole region 
while $u_{\rm eff}^{(2)}$ changes sign at $R\approx 2$ fm. 

\begin{figure}[hbt]
\begin{center}
\includegraphics[width=10cm]{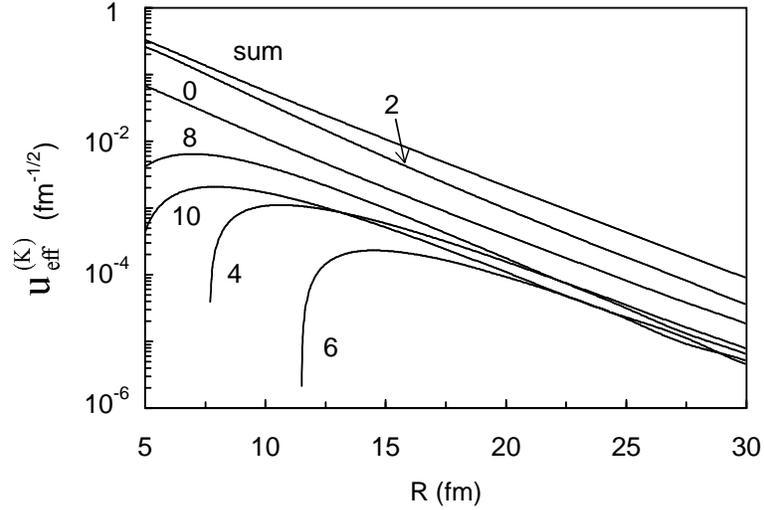}
\end{center}
\caption{Same as in Fig.~\ref{Fig3} in logarithmic scale.
\label{Fig4}}
\end{figure}

Since all other components are not visible in Fig.~\ref{Fig3}, 
we turn in Fig.~\ref{Fig4} to a logarithmic scale. 
For relative distances from $R=6$ fm up to 25 fm, 
the contribution of the $K=8$ component is larger than the contributions of the $K=4$ and 6 components. 
This is due to a zero around 10 fm in both components. 

The curves in Figs.~\ref{Fig3} and \ref{Fig4} indicate that the product 
$u_E(R) u_{\rm eff}^{(K)}(R)$ for $K=0$ changes sign several times. 
The integral $I_E^{(0)}(R)$ is first positive, starts decreasing at the first zero of $u_E$, 
changes sign near 2 fm and increases again at the second zero of $u_E$. 
The combined effect of both zeros results in a cancellation between the internal and external parts 
of the corresponding limit $I_E^{(0)}(+\infty)$. 
These zeros at short distances are due to the existence of two bound states in the potential 
(one physical and one forbidden). 
The numbers and locations of zeros are typical of the $^6$He $\beta$ decay 
so that the cancellation should not occur for the decay of other halo nuclei. 
It would not occur so strongly with a single zero. 

For $K = 2$, the combined effects of the zero of $u_{\rm eff}^{(2)}$ and of the first zero of $u_E$ 
is just a small plateau near 2 fm in Fig.~\ref{Fig2}. 
The second zero of $u_E$ gives a minimum near 5 fm. 
The resulting $I_E^{(2)}(+\infty)$ for the $K=2$ component also yields an important cancellation, 
but not as strong as in the $K=0$ case. 

The effective functions for $K=4$ and 6 are very small in the region where $u_E$ 
keeps a constant sign and lead to negligible contributions. 
Let us recall here that the convergence of low-$K$ components is not reached 
if $K_{\max}$ is not large enough \cite{DDB03}. 

A new situation appears for the $K=8$ component. 
The effective wave function is much smaller than $K=0$ or 2 but the cancellation is minimal 
since it does not change sign. 
Hence it gives the second largest $I_E^{(K)}$ at infinity. 
The same mechanism applies for the smaller $K \ge 10$ components. 
The $K=10$ integral still contributes significantly to the total sum. 

\begin{figure}[thb]
\begin{center}
\includegraphics[width=10cm] {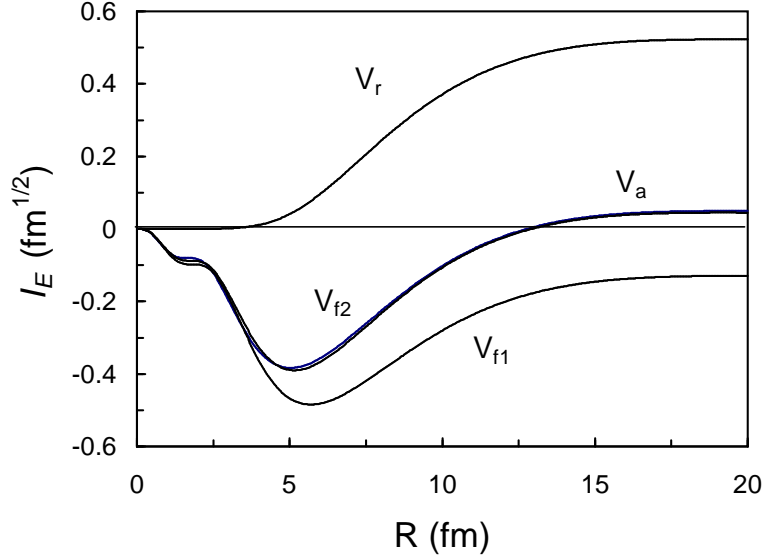}
\end{center}
\caption{Integrals $I_E (R) = \sum_K I_E^{(K)} (R)$ at $E=1$ MeV 
calculated with $K_{\rm max}=24$ and $R_{\rm max}=30$ fm for different $\alpha+d$ potentials. 
\label{Fig5}}
\end{figure}

In Fig.~\ref{Fig5}, the integrals $I_E (R) = \sum_K I_E^{(K)} (R)$ calculated 
at the energy $E=1$ MeV are represented for the different potentials. 
The repulsive potential $V_r$ displays a strongly different behavior from the other potentials. 
At this energy, $V_a$ and $V_{f2}$ give almost the same result. 
This is due to the fact that, because of their similar phase shifts, 
their node near 5 fm in the scattering wave $u_E$ is at nearly the same location. 
The comparison with $V_{f1}$, where this node is about 1 fm farther away and leads therefore 
to weaker cancellation effects, shows the major role played by this node. 

The $K=2$ and $K=8$ components of the three-body hyperspherical wave function 
of the $^6$He nucleus give dominant contributions to the integral $I_E(R)$ at large values of $R$ 
and thus to the Gamow-Teller reduced transition probability $B_{\rm GT}(E)$. 
This finding contradicts the assumption in Ref.~\cite{zhu93}, 
that the $K=0$ and 2 dominant contributions to the energy are sufficient 
to study this $\beta$ decay mode. 

\subsection{Transition probability per time and energy units}

\begin{figure}[htb]
\begin{center}
\includegraphics[width=10cm]{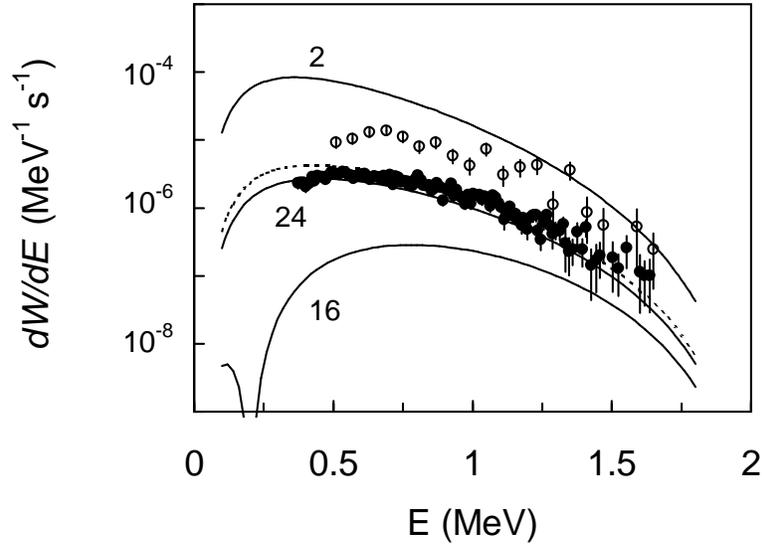}
\end{center}
\caption{Transition probability per time and energy units $dW/dE$ 
of the $^6$He $\beta$ decay into the $\alpha+d$ continuum calculated 
with the $\alpha+d$ potential $V_a$ and $R_{\rm max}=30$ fm for several values of $K_{\rm max}$. 
Experimental data are from Refs.~\cite{bor93} (open dots) and \cite{ABB02} (full dots). 
The dotted line corresponds to the $\alpha + N$ potential of
Ref.~\protect\cite{KKN79} for the $^6$He wave function.
\label{Fig6}}
\end{figure}

Before comparing transition probabilities with experiment, 
we discuss convergence aspects. 
To study the convergence with respect to the value of the maximal hypermomentum $K_{\rm max}$, 
we display in Fig.~\ref{Fig6} the transition probabilities [Eq.~(\ref{eq310})] 
calculated with potential $V_a$ for $K_{\rm max}=2$, 16, and 24. 
In each case the $\alpha+N$ interaction has been renormalized to reproduce the $^6$He binding energy.
In a logarithmic scale, the results for $K_{\rm max}=22$ are essentially identical to those for $K_{\rm max}=24$. 
From Fig.~\ref{Fig6}, we can see that using large $K_{\rm max}$ values is crucial to obtain a good accuracy. 
This is analyzed in more detail below.

Additionally, the convergence is faster for the repulsive potential $V_r$ and 
for the folding potentials $V_{f1}$ and $V_{f2}$. 
In the case of $V_r$, the transition probabilities for $K_{\rm max}=16$ and $K_{\rm max}=24$ show 
almost the same features but have a larger value than for $V_a$. 
However, even in this case, the choice $K_{\rm max}=2$ in Ref.~\cite{zhu93} is not realistic. 

A calculation with $K_{\rm max}=24$ performed with the original wave function 
of Ref.~\cite{DDB03}, i.e. with the $\alpha + n$ potential of Kanada et al \cite{KKN79}, 
and the $\alpha + d$ potential $V_a$ is displayed as a dotted line. 
The results are somewhat larger and in less good agreement with experiment at low energies 
but they nicely reproduce both the shape and order of magnitude of the data. 
This indicates that the present model is not much sensitive to details of the 
model describing the $^6$He wave function and confirms that convergence and an appropriate 
$\alpha + d$ potential are the crucial elements. 
 
\begin{table}[h]
\caption{Components $I_E^{(K)} (\infty)$ as a function of $K$ and $K_{\rm max}$ at $E = 1$ MeV}
\begin{tabular}{rrrrrrr}
\hline
$K$ & $K_{\rm max}=2$ & $K_{\rm max}=16$ & $K_{\rm max}=18$ & $K_{\rm max}=20$ & $K_{\rm max}=22$ & $K_{\rm max}=24$\\
\hline
$0$ & $ 3.67\times 10^{-2}$ & $ 2.84\times 10^{-3}$ & $ 4.02\times 10^{-3}$ & $ 5.21\times 10^{-3}$ & $ 5.43\times 10^{-3}$ & $ 5.74\times 10^{-3}$\\
$2$ & $ 1.49\times 10^{-1}$ & $-2.58\times 10^{-2}$ & $-2.09\times 10^{-2}$ & $-1.34\times 10^{-2}$ & $-1.17\times 10^{-2}$ & $-9.45\times 10^{-3}$\\
$4$ &  & $ 4.99\times 10^{-3}$ & $ 5.27\times 10^{-3}$ & $ 4.75\times 10^{-3}$ & $ 4.90\times 10^{-3}$ & $ 4.78\times 10^{-3}$\\
$6$ &  & $ 1.14\times 10^{-3}$ & $ 1.13\times 10^{-3}$ & $ 1.56\times 10^{-4}$ & $ 5.12\times 10^{-5}$ & $-3.06\times 10^{-4}$\\
$8$ &  & $ 2.58\times 10^{-2}$ & $ 2.66\times 10^{-2}$ & $ 2.79\times 10^{-2}$ & $ 2.84\times 10^{-2}$ & $ 2.90\times 10^{-2}$\\
$10$ &  & $ 8.64\times 10^{-3}$ & $ 8.90\times 10^{-3}$ & $ 9.37\times 10^{-3}$ & $ 9.55\times 10^{-3}$ & $ 9.78\times 10^{-3}$\\
$12$ &  & $ 4.32\times 10^{-4}$ & $ 4.15\times 10^{-4}$ & $ 7.32\times 10^{-5}$ & $ 8.38\times 10^{-6}$ & $-1.50\times 10^{-4}$\\
$14$ &  & $ 2.03\times 10^{-3}$ & $ 2.14\times 10^{-3}$ & $ 2.21\times 10^{-3}$ & $ 2.26\times 10^{-3}$ & $ 2.30\times 10^{-3}$\\
$16$ &  & $ 2.01\times 10^{-3}$ & $ 2.11\times 10^{-3}$ & $ 2.33\times 10^{-3}$ & $ 2.42\times 10^{-3}$ & $ 2.53\times 10^{-3}$\\
$18$ &  &  & $ 4.82\times 10^{-4}$ & $ 4.70\times 10^{-4}$ & $ 4.71\times 10^{-4}$ & $ 4.62\times 10^{-4}$\\
$20$ &  &  &  & $ 1.49\times 10^{-4}$ & $ 1.45\times 10^{-4}$ & $ 1.22\times 10^{-4}$\\
$22$ &  &  &  &  & $ 3.26\times 10^{-4}$ & $ 3.49\times 10^{-4}$\\
$24$ &  &  &  &  &  & $ 1.59\times 10^{-4}$\\
\hline
sum & $ 1.86\times 10^{-1}$ & $ 2.21\times 10^{-2}$ & $ 3.01\times 10^{-2}$ & $ 3.92\times 10^{-2}$ & $ 4.23\times 10^{-2}$ & $ 4.53\times 10^{-2}$\\
\hline
\end{tabular}
\label{table0}
\end{table}

The low value of $dW/dE$ results from different cancellation effects 
which themselves are sensitive to the convergence of the different components 
of the three-body wave function. 
The final order of magnitude implies an accurate treatment of the convergence 
of the wave function and in particular of its halo part. 
In order to illustrate this mechanism, we display in Table \ref{table0} 
the numerical values of the components $I_E^{(K)} (\infty)$ of the Gamow-Teller integral 
for different values of the parameter $K_{\rm max}$ at 1 MeV. 
Let us start with $K_{\rm max} \ge 16$. 
For each value of $K_{\rm max}$, one observes that the dominant components 
are indeed $K=2$ and $K=8$. 
Components beyond $K = 16$ become rather small. 
However, when $K_{\rm max}$ is increased, all components $I_E^{(K)} (\infty)$ 
of the matrix element are modified. 
As emphasized in Ref.~\cite{DDB03}, increasing $K_{\rm max}$ in the three-body model 
does not only mean adding components but, even more, improving the convergence 
of lower $K$ components. 
This is illustrated when following a row in Table \ref{table0}. 
The value of each component slowly converges when $K_{\rm max}$ is increased. 
If the experimental data were much more accurate, higher values of $K_{\rm max}$ 
should probably be considered. 

The dominant $K = 2$ and $K = 8$ components have opposite signs. 
This effect adds another level of cancellation in the matrix element. 
It increases the role of the other components and especially 
the collective role of high-$K$ components. 
Finally a comparison with the first column shows why a calculation 
restricted to $K_{\rm max} = 2$ has little meaning: 
the $K = 2$ component is too large, has a wrong sign, and is not counterbalanced 
by other components. 
 
\begin{figure}[htb]
\begin{center}
\includegraphics[width=10cm]{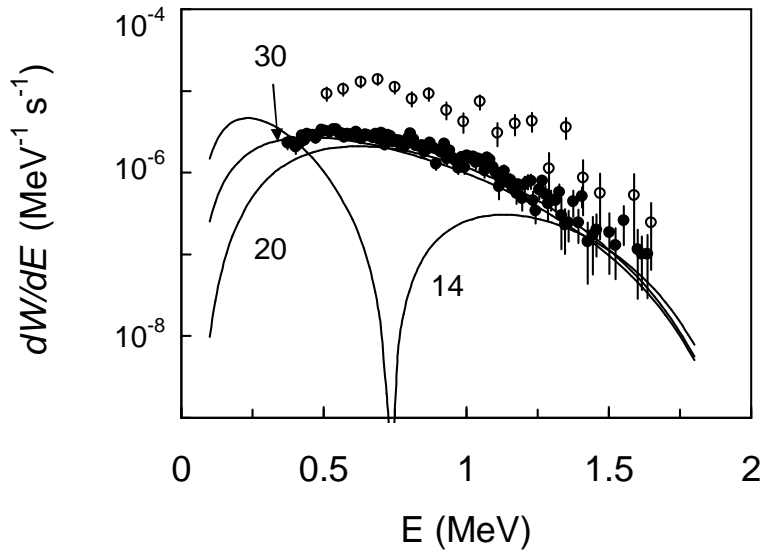}
\end{center}
\caption{Transition probability per time and energy units $dW/dE$ of the $^6$He $\beta$ decay 
into the $\alpha+d$ continuum calculated 
with the $\alpha+d$ potential $V_a$ and $K_{\rm max}=24$ for several values of $R_{\rm max}$. 
Experimental data are from Refs.~\cite{bor93} (open dots) and \cite{ABB02} (full dots).
\label{Fig7}}
\end{figure}

In Fig.~\ref{Fig7}, the transition probability $dW/dE$ obtained with potential $V_a$ 
for a fixed $K_{\rm max}=24$ is presented for different values of $R_{\rm max}$, 
i.e., 14 fm, 20 fm, and 30 fm. 
Calculations for $R_{\rm max}=35$ fm show that convergent results are obtained 
at $R_{\rm max} \approx 30$ fm. 
Taking properly account of the halo extension is very important in a correct treatment 
of the very small transition probability $dW/dE$ of the $^6$He $\beta$ decay into $\alpha+d$. 

\begin{figure}[thb]
\begin{center}
\includegraphics[width=10cm]{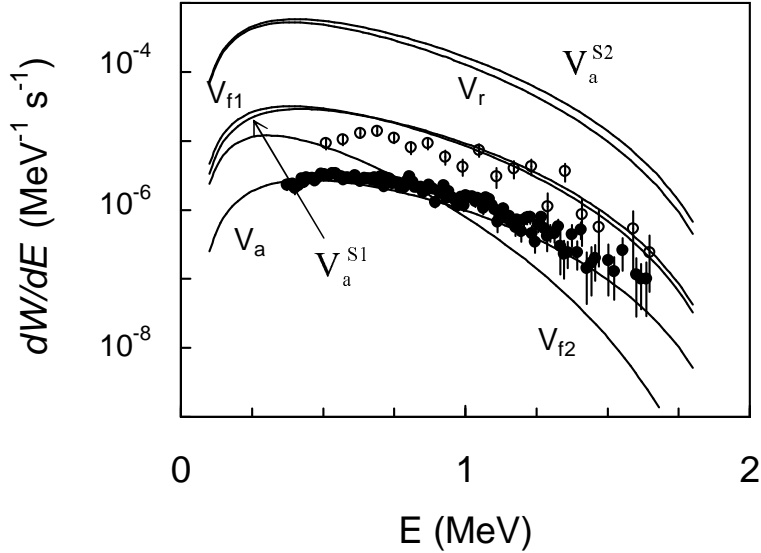}
\end{center}
\caption{Transition probability per time and energy units $dW/dE$ of the $^6$He $\beta$ decay 
into the $\alpha+d$ continuum calculated with different $\alpha+d$ potentials 
for $K_{\rm max}=24$ and $R_{\rm max}=30$ fm. 
Experimental data are from Refs.~\cite{bor93} (open dots) and \cite{ABB02} (full dots).
\label{Fig8}}
\end{figure}

In Fig.~\ref{Fig8}, we display the transition probability for different potentials, 
calculated with $K_{\rm max}=24$ and $R_{\rm max}=30$ fm. 
The best description of the experimental data of Ref.~\cite{ABB02} 
is obtained with the attractive potential $V_a$.  
The worst results correspond to the repulsive potential $V_r$, which has no bound state 
and for which the description of the $s$-wave phase shift at low energies is poor. 
With potential $V_r$, the location of the nodes in the scattering wave 
does not lead to a cancellation (see Fig.~\ref{Fig5}). 
The folding potentials $V_{f1}$ and $V_{f2}$ have intermediate behaviors. 
Potential $V_{f1}$ overestimates the recent data while potential $V_{f2}$ provides 
a better order of magnitude but its energy dependence disagrees with the experimental one. 

We have also performed a calculation with the original wave function of Ref.~\cite{DDB03} 
and with a folding potential based on the $\alpha + n$ potential of Kanada et al \cite{KKN79} 
of $V_{f2}$ type, i.e. multiplied by 1.30 in order to fairly reproduce the phase shift. 
The results are indistinguishable from the curve labeled $V_{f2}$ at the scale of the figure. 
The shape of this curve is thus mostly due to the node locations of the $\alpha + d$ 
wave function. 

The success of the deep Gaussian potential could be attributed to the fact that it simultaneously 
reproduces both the $^6$Li ground state binding energy and the $s$-wave phase shift at low energies. 
However the discussion of Figs.~\ref{Fig2}-\ref{Fig4} indicates that an important 
ingredient is the existence of two nodes in $u_{\rm eff}$. 
In order to test this assumption, we remove the non-physical ground state of $V_a$ by using a pair 
of supersymmetric transformations \cite{Ba-87}. 
The resulting phase-equivalent potential $V_a^{S1}$ has exactly the same $^6$Li ground-state energy 
and the same $s$-wave phase shift as $V_a$ but its scattering wave functions have one node less 
at small distances. 
The resulting $dW/dE$ is about one order of magnitude larger and resembles the one obtained 
with the folding potential $V_{f1}$ (see Fig.~\ref{Fig8}). 
Notice however that $V_{f1}$ has two bound states but does not well reproduce the phase shifts. 
A second phase-equivalent potential $V_a^{S2}$ is obtained by removing the $^6$Li ground state 
from $V_a^{S1}$ with another pair of transformations. 
This repulsive potential has still exactly the same phase shifts as $V_a$ but no bound state. 
Its scattering wave functions have no node near the origin. 
As expected, the corresponding transition probability $dW/dE$ is now very close to the one 
obtained with potential $V_r$. 
The comparison emphasizes the crucial role played by the forbidden bound state, 
in addition to the physical $^6$Li ground state, for reproducing the order 
of magnitude of the experimental data. 

The total transition probabilities for different potentials are given in the first row 
of Table \ref{table1}. 
The second row contains results corresponding to the experimental cutoff \cite{ABB02}. 
The values in the last columns are derived from experimental branching ratios 
and from the $^6$He half life \cite{ABB02}. 
As expected from the previous discussion, the result obtained with the Gaussian potential $V_a$ 
falls within the experimental error bars of Ref.~\cite{ABB02}. 
The other results are too large, especially with the repulsive potential. 
\begin{table}[h]
\caption{Total transition probability per second $W$ (in $10^{-6}$ s$^{-1}$)
for the $\beta$ decay of $^6$He into $\alpha+d$.}
\begin{tabular}{ccccccc} 
\hline
    & $V_a$  &  $V_{f1}$  & $V_{f2}$  & $V_r$  & Exp.~\cite{bor93} & Exp.~\cite{ABB02} \\ 
\hline
$E>0$ MeV   & $1.95$ & $25.1$ & $5.80$ & $340$ &                   & $2.2\pm 1.1$      \\
$E>0.37$ MeV& $1.54$ & $18.2$ & $3.32$ & $246$ & $7.6\pm 0.6$      & $1.5\pm 0.8$      \\ 
\hline
\end{tabular}
\label{table1}
\end{table}
\section{A comment on $R$-matrix fits}
The $R$-matrix method has been extended by Barker \cite{bar94} to the $\beta$ delayed 
deuteron emission. 
It has been applied to analyze recent experimental results \cite{ABB02}. 
Like in other models, it is crucial in the $R$-matrix method to take care 
of the large extension of the halo. 
Without entering into details which are explained in Refs.~\cite{bar94,ABB02}, 
this is achieved by introducing external corrections proportional to the integral 
\beq
\mathcal{I}_E (a) = \int_a^{\infty} \frac{u_i(R)}{u_i(a)} u_E (R) dR,
\eeqn{eq401}
where $a$ is the $R$-matrix channel radius 
and $u_E$ is replaced by its asymptotic expression (\ref{eq220}). 
The factor $u_i(a)$ eliminates the problem of normalizing 
the approximation for the initial wave function $u_i(R)$. 
In Ref.~\cite{bar94}, the notation $E_i(R)$ is used for $u_i(R)/u_i(a)$. 

In the model of Ref.~\cite{bar94}, the asymptotic form of the two-body $\alpha+$dineutron 
system is employed for $u_i$, 
\beq
u_i^{\alpha+2n} (R) = \exp [ - (2\mu_{(12)3} m_N |E_B|/\hbar^2)^{1/2} R],
\eeqn{eq402}
where $E_B = -0.975$ MeV. 
However, three-body asymptotics are rather different from this expression. 
In Eq.~(\ref{eq104}), this role is played by the effective radial wave function $u_{\rm eff}$ 
defined by Eq.~(\ref{eq305}). 
In order to avoid the knowledge of three-body wave functions, 
we suggest here an expression, 
\beq
u_i^{\alpha+n+n} (R) = R \int_0^{\infty} \rho^{-5/2} \exp [ - (2m_N|E_B|/\hbar^2)^{1/2} \rho] r u_d(r) dr, 
\eeqn{eq403}
which is the projection of three-body asymptotics \cite{Me-74,DDB03} on the deuteron wave function. 
For a pointlike deuteron described with $u_d(r) \propto r^{-1}\delta (r)$, this function becomes 
\beq
u_i^{\alpha+2n,{\rm cor.}} (R) = R^{-3/2} \exp [ - (2\mu_{(12)3} m_N |E_B|/\hbar^2)^{1/2} R].
\eeqn{eq404}
It differs from Eq.~(\ref{eq402}) by the power factor $R^{-3/2}$. 
This simple expression also deserves being evaluated. 

In Table \ref{table2}, we present 
the integral (\ref{eq401}) calculated with $u_i^{\alpha+2n}$, $u_i^{\alpha+n+n}$, 
$u_i^{\alpha+2n,{\rm cor.}}$, and $u_{\rm eff}$ at two typical energies 
for different values of the channel radius $a$. 
\begin{table}[ht]
\caption{External integrals [Eq.~(\ref{eq401})] in $R$-matrix fits: 
$\mathcal{I}_E^{\alpha+2n}$, $\mathcal{I}_E^{\alpha+2n,{\rm cor.}}$, 
$\mathcal{I}_E^{\alpha+n+n}$, and $\mathcal{I}_E^{\rm eff}$ are 
calculated with $u_i^{\alpha+2n}$ [Eq.~(\ref{eq402})], $u_i^{\alpha+2n,{\rm cor.}}$ [Eq.~(\ref{eq404})], 
$u_i^{\alpha+n+n}$ [Eq.~(\ref{eq403})], and $u_{\rm eff}$ [Eq.~(\ref{eq305})], respectively,  
for different values of the channel radius $a$ (in fm) and the relative energy $E$ (in MeV).}
\begin{tabular}{ccccccc} 
\hline
$E$ &  $a$ & $\mathcal{I}_E^{\alpha+2n}$ & $\mathcal{I}_E^{\alpha+2n,{\rm cor.}}$ 
& $\mathcal{I}_E^{\alpha+n+n}$ & $\mathcal{I}_E^{\rm eff}$ \\ 
\hline
0.5 & 4.0 & 1.162 & 0.224 & 0.470 & 0.598 \\
    & 4.5 & 1.391 & 0.384 & 0.651 & 0.751 \\
    & 5.0 & 1.616 & 0.552 & 0.834 & 0.913 \\
    & 5.5 & 1.834 & 0.726 & 1.018 & 1.082 \\
    & 6.0 & 2.046 & 0.903 & 1.201 & 1.432 \\
\hline
1.0 & 4.0 & 1.464 & 0.367 & 0.698 & 0.882 \\
    & 4.5 & 1.754 & 0.601 & 0.952 & 1.091 \\
    & 5.0 & 2.024 & 0.839 & 1.200 & 1.307 \\
    & 5.5 & 2.172 & 1.075 & 1.437 & 1.522 \\
    & 6.0 & 2.496 & 1.303 & 1.660 & 1.732 \\
\hline
\end{tabular}
\label{table2}
\end{table}
One observes that the results obtained with the two-body asymptotic expression (\ref{eq402}) 
are rather far from the realistic values obtained with $u_{\rm eff}$, even for $a = 6$ fm. 
A much better approximation is given by the three-body asymptotic expression (\ref{eq403}), 
especially at higher relative energies. 
The corrected two-body approximation is smaller than the three-body approximation 
and not really close to the reference results. 

\begin{figure}[htb]
\begin{center}
\includegraphics[width=10cm]{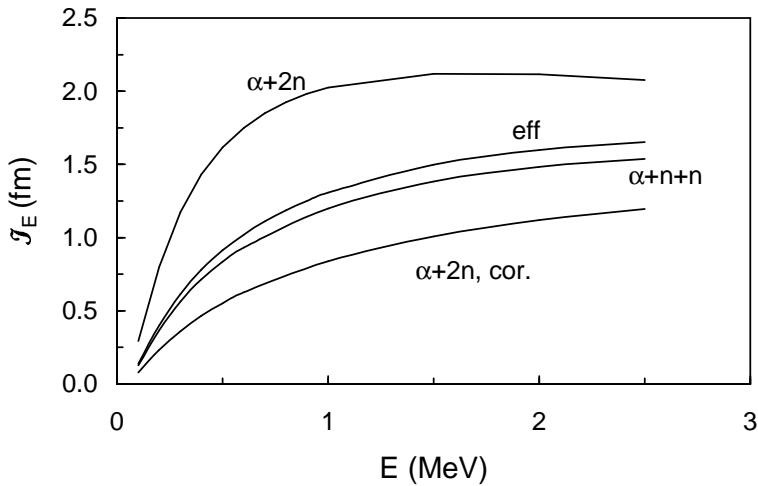}
\end{center}
\caption{External integrals $\mathcal{I}_E^{\alpha+2n}$, $\mathcal{I}_E^{\alpha+2n,{\rm cor.}}$, 
$\mathcal{I}_E^{\alpha+n+n}$, and $\mathcal{I}_E^{\rm eff}$ 
as a function of the relative energy $E$ for $a = 5$ fm. 
\label{Fig9}}
\end{figure}

In $R$-matrix theory however, because parameters are fitted,
the absolute normalization of the integrals displayed in Table \ref{table2} is not important. 
It can be absorbed in a renormalization of the constants. 
The crucial property is their energy dependence. 
The four types of external integrals are shown in Fig.~\ref{Fig9} 
as a function of energy for the typical value $a = 5$ fm. 
One observes that $\mathcal{I}_E^{\alpha+2n,{\rm cor.}}$, 
$\mathcal{I}_E^{\alpha+n+n}$, and $\mathcal{I}_E^{\rm eff}$ have very similar energy dependences. 
On the contrary, $\mathcal{I}_E^{\alpha+2n}$ displays a different shape. 
Hence, using this approximation in $R$-matrix fits may significantly distort 
the energy shape of the $\beta$ delayed deuteron spectrum. 
The $\alpha+n+n$ approximation offers a very good approximation of the model results. 
Nevertheless, the corrected two-body expression provides 
the simplest significant improvement for $R$-matrix calculations. 
\section{Conclusions}
In the present work, we studied the $\beta$-decay process of the $^6$He halo nucleus 
into the $\alpha+d$ continuum in the framework of a three-body model. 
Three-body hyperspherical bound-state wave functions on a Lagrange mesh and two-body  $\alpha+d$ 
scattering wave functions have been used. 
For the calculation of the $\beta$-decay transition probabilities per time and energy units, 
several $\alpha+d$ potentials were tested: an attractive Gaussian potential \cite{dub94} 
with a deep forbidden bound state, folding potentials derived from the $\alpha+N$ $p$-wave potential 
of Ref.~\cite{vor95}, and a repulsive potential \cite{zhu93}. 

We confirm that the low experimental values result from a strong cancellation in the Gamow-Teller 
matrix element describing the transition to the continuum \cite{BSD94}. 
This cancellation occurs between the internal and halo parts of the matrix element 
and is thus very sensitive to the halo description. 
Reaching convergence is not easy: 
the two-body and three-body wave functions must extend up to about 30 fm. 
From the analysis of the theoretical results, we have found that converged results 
require large values of the maximal hypermomentum $K_{\rm max}$. 
The dominant contributions to the transition probability come from the $K=2$, $K=8$, and $K=10$ components 
of the three-body hyperspherical wave function. 
The $K=0$ contribution is small due to an almost perfect cancellation of its internal 
and external parts in the Gamow-Teller matrix element. 
The $K=2$ and $K=8$ components have opposite signs which enhances the importance 
of other, and especially high-$K$, components. 

The experimental transition probabilities per time and energy units of Ref.~\cite{ABB02} 
are well described with the deep Gaussian potential of Ref.~\cite{dub94} 
which fairly reproduces the $^6$Li binding energy and the $\alpha+d$ $s$-wave phase shifts. 
The quality of the agreement arises from the node structure of the initial and final wave functions 
in the internal part. 
With the help of phase-equivalent potentials derived with supersymmetric transformations, 
we have shown that the role of the forbidden state is also essential. 
We realize that the efficiency of the deep potential may be somewhat fortuitous 
since the nodes of the scattering wave function have to be at very precise locations. 
The fact that the data can be reproduced does not mean that the present 
model or the simple Gaussian potential are perfect. 
However the reasonable agreement with the data obtained with the same $\alpha+d$ potential 
but another $^6$He wave function indicates that the present model interpretation should be trustable. 
Most importantly the existence of a good agreement with experiment points toward the ingredients 
that are crucial in the interpretation of the $\beta$ delayed deuteron decay of $^6$He. 
One can expect a completely different behavior for this $\beta$-decay mode in the case of $^{11}$Li. 
Indeed the $^6$He-case cancellations require precise numbers of nodes and precise locations of these nodes 
and it is very unlikely that this could occur so perfectly in another case. 

Our results allow testing the validity of external corrections necessary in the $R$-matrix method \cite{bar94}. 
We have shown that, in order to reduce a systematic bias in the integrals over the external region, 
the two-body asymptotics can usefully be replaced by three-body asymptotics 
or, more simply, be corrected by a factor $R^{-3/2}$. 

Further progress on this $\beta$-decay mode must come from consistent fully microscopic descriptions 
of the bound and scattering states. 
The results obtained with a microscopic cluster model \cite{cso94} still agree qualitatively 
with the most recent data \cite{ABB02} but overestimate them by about a factor of two. 
Progress may be expected from the possibility of calculating $^6$He wave functions ab initio \cite{PWC04} 
from realistic two- and three-body forces. 
However the present study shows that a successfull description of the $\beta$ delayed deuteron emission 
will require very accurate bound-state wave functions up to distances as large as 30 fm 
and a development of consistent scattering wave functions. 
The accidental cancellation occurring in this process will make a fully ab initio description 
particularly difficult. 
\section*{Acknowledgments}
This text presents research results of the Belgian program P5/07 on 
interuniversity attraction poles initiated by the Belgian-state 
Federal Services for Scientific, Technical and Cultural Affairs (FSTC).
P.D. and E.M.T. acknowledge the support of the National Fund for Scientific Research (FNRS), Belgium. 
E.M.T thanks the PNTPM group of ULB for its kind hospitality during his stay in Brussels. 
\end{document}